\documentclass[journal]{IEEEtran}

\usepackage{mathrsfs}
\usepackage[noadjust]{cite}
\usepackage{graphicx,color,overpic,psfrag}
\usepackage{amsmath, amssymb}
\usepackage{latexsym}
\usepackage{bm}
\usepackage{amssymb}
\usepackage{cases}
\usepackage{array}
\usepackage{fancyhdr}
\usepackage{setspace}

\ifCLASSOPTIONcompsoc
\usepackage[caption=false,font=normalsize,labelfont=sf,textfont=sf]{subfig}
\else
\usepackage[caption=false,font=footnotesize]{subfig}
\fi

\usepackage{url}
\usepackage{algpseudocode}
\usepackage{algorithm}
\usepackage{blkarray}
\usepackage{booktabs}

\usepackage{multirow}
\usepackage{dsfont}
\usepackage{tabularx}
\usepackage[table]{xcolor}

\usepackage{amsfonts}

\usepackage{letltxmacro}

\graphicspath{{figure/}}

\newtheorem{remark}{Remark}

\newtheorem{prop}{Proposition}

\newcommand{\figref}[1]{Fig. \ref{#1}}

\newcommand{\alref}[1]{\textbf{Algorithm \ref{#1}}}

\newcommand{\propref}[1]{\emph{Proposition \ref{#1}}}

\newcommand{\Exp}{{\mathsf{E}}}
\newcommand{\expect}[1]{\Exp\left\{#1\right\}}

\newcommand{\tr}[1]{\mathsf{tr}\left\{#1\right\}}
\newcommand{\diag}[1]{\mathsf{diag}\left\{#1\right\}}

\newcommand{\cK}{\mathcal{K}}
\newcommand{\cL}{\mathcal{L}}

\newcommand{\cO}{\mathcal{O}}
\newcommand{\cP}{\mathcal{P}}

\newcommand{\bn}{\mathbf{n}}

\newcommand{\bu}{\mathbf{u}}
\newcommand{\bv}{\mathbf{v}}

\newcommand{\bx}{\mathbf{x}}
\newcommand{\by}{\mathbf{y}}

\newcommand{\bH}{\mathbf{H}}
\newcommand{\bI}{\mathbf{I}}

\newcommand{\bK}{\mathbf{K}}

\newcommand{\bQ}{\mathbf{Q}}
\newcommand{\bR}{\mathbf{R}}

\newcommand{\bU}{\mathbf{U}}
\newcommand{\bV}{\mathbf{V}}

\newcommand{\bX}{\mathbf{X}}

\newcommand{\bzero}{\mathbf{0}}

\newcommand{\bSigma}{{\boldsymbol\Sigma}}

\newcommand{\bLambda}{{\boldsymbol\Lambda}}

\newcommand{\bGamma}{{\boldsymbol\Gamma}}
\newcommand{\bgamma}{{\boldsymbol\gamma}}
\newcommand{\bPsi}{{\boldsymbol\Psi}}
\newcommand{\bpsi}{{\boldsymbol\psi}}

\newcommand{\bOmega}{{\boldsymbol\Omega}}

\newcommand{\ntb}{\notag\\}

\newcommand{\R}{\mathbb{R}}
\newcommand{\C}{\mathbb{C}}

\newcommand{\SE}{\mathrm{SE}}

\newcommand{\EE}{\mathrm{EE}}
\newcommand{\SAR}{\mathrm{SAR}}
\newcommand{\SEde}{\overline{\SE}}
\newcommand{\EEde}{\overline{\EE}}
\newcommand{\Pck}{P_{\mathrm{c},k}}

\newcommand{\s}{\frac{1}{\sigma ^{2}}}

\newcommand{\opt}{\mathrm{opt}}

\newcommand{\beamH}{\widetilde{\bH}}

\allowdisplaybreaks

\begin{document}

\title{Energy-Efficient Precoding in Electromagnetic Exposure-Constrained Uplink Multiuser MIMO}

\author{Jiayuan~Xiong, Li~You, Derrick~Wing~Kwan~Ng, Wenjin~Wang, and~Xiqi~Gao

\thanks{Copyright (c) 2015 IEEE. Personal use of this material is permitted. However, permission to use this material for any other purposes must be obtained from the IEEE by sending a request to pubs-permissions@ieee.org.}%
\thanks{
Jiayuan Xiong, Li You, Wenjin Wang, and Xiqi Gao are with the National Mobile Communications Research Laboratory, Southeast University, Nanjing 210096, China, and also with the Purple
Mountain Laboratories, Nanjing 211100, China (e-mail: jyxiong@seu.edu.cn; liyou@seu.edu.cn; wangwj@seu.edu.cn; xqgao@seu.edu.cn). \textit{(Corresponding author: Li You.)}
}
\thanks{
Derrick Wing Kwan Ng is with the School of Electrical Engineering and Telecommunications, University of New South Wales, NSW 2052, Australia (e-mail: w.k.ng@unsw.edu.au).
}
}

\maketitle
\begin{abstract}
User electromagnetic (EM) exposure is continuously being exacerbated by the evolution of multi-antenna portable devices. To mitigate the effects of EM radiation, portable devices must satisfy tight regulations on user exposure level, generally measured by specific absorption rate (SAR). To this end, we investigate the SAR-aware uplink precoder design for the energy efficiency (EE) maximization in multiuser multiple-input multiple-output transmission exploiting statistical channel state information (CSI). As the objective function of the design problem is computationally demanding in the absence of closed form, we present an asymptotic approximation of the objective to facilitate the precoder design. An iterative algorithm based on Dinkelbach's method and sequential optimization is proposed to obtain an optimal solution of the asymptotic EE optimization problem. Based on the transformed problem, an iterative SAR-aware water-filing scheme is further conceived for the EE optimization precoding design with statistical CSI. Numerical results illustrate substantial performance improvements provided by our proposed SAR-aware energy-efficient transmission scheme over the traditional baseline schemes.
\end{abstract}



\section{Introduction}
The prevalence of portable and wearable devices for wireless communications in the past decades, e.g., laptops, smart-watches, and smart-phones, has made extensive influences on our everyday lives. These devices, as well as the related applications, have generated vast improvements to, e.g., networking, communications, and entertainment. However, these devices produce electromagnetic (EM) signals when they operate, thus emitting hazardous EM radiation and posing potential health threats to neighbouring users during use \cite{castellanos2019dynamic}. Therefore, all wireless devices must adhere to certain regulations on EM exposure limits. In fact, specific absorption rate (SAR) is a universally accepted metric for user exposure measurement \cite{lin2001closed}. More specifically, SAR, with a unit of W/kg, measures the energy absorbed by per unit mass of human tissue. In practical commercial scenarios, wireless portable devices must satisfy the SAR constraints enforced by regulatory agencies around the world. For instance, the maximum tolerable SAR on partial-body exposure is limited to $1.6$ W/kg in the US \cite{federal2001evaluating}.

In practice, the worst-case SAR compliance for single-antenna devices can be satisfied via simply reducing the transmit power at the expense of low system data rates \cite{ying2017sum}. In nowadays and future communication systems, multi-antenna devices have become the mainstream to enable high spectrum efficiency (SE). However, for these devices, it is much more challenging to handle the worst-case EM exposure due to huge variations of SAR when multiple antennas are equipped at the transmitter \cite{Chim2004Investigating,zhang2020specific}. Moreover, state-of-the-art cellular devices usually support more than one standard, e.g., LTE, WiFi, and Bluetooth, which hinders the systems concurrently in satisfying the SAR constraints. Since commercial devices have become more sophisticated and powerful due to rapid evolution of wireless transmission technologies \cite{ying2015closed}, uplink user EM exposure is getting worse and worse. In addition, portable devices are usually more close to human beings than base station (BSs). Hence, uplink SAR-aware signaling techniques have become increasingly important for future cellular devices to satisfy EM exposure limits, especially in multiple-input multiple-output (MIMO) systems \cite{castellanos2019dynamic,ying2017sum,zhang2020specific}.

Extensive works on SAR-aware transmission schemes have been emerging in the last few years. For example, in \cite{Hochwald2012minimizing}, an SAR-aware design of transmit signals was first investigated in SAR-constrained MIMO uplink systems and a quadratic model for the SAR metric was proposed. In \cite{zhang2017analysis}, the authors analyzed the MIMO-SAR performances of six dual-antenna handsets and simplified the MIMO-SAR measurement. In \cite{Heliot2020exposure}, the authors provided a more practical two-antenna SAR model based on the assumption that each portable device was equipped with two transmit antennas. In \cite{Ying2013transmit}, an optimal precoding strategy for the SE maximization was designed under one SAR constraint. In \cite{ying2015closed}, a more general SAR-aware precoder design incorporating various SAR constraints was investigated to maximize the system capacity. Note that available works on SAR-aware analyses were mainly dedicated to improving the system SE. Nevertheless, energy efficiency (EE), as another vital and considerable system performance, is not given much attention in SAR-aware transmission designs. In addition, acquiring perfect instantaneous channel state information (CSI) is an arduous and expensive job in many practical scenarios. Furthermore, in fast time-varying channels of fast-moving user terminals (UTs), the obtained instantaneous CSI becomes outdated easily. Therefore, as a compromise approach, statistical CSI has been exploited in some research activities \cite{Wen11On,Gao09Statistical,You2020network,you2020reconfigurable,li2020ffr,You2020RIStradeoff} due to its relatively easy availability. However, these existing works focused on the conventional transmission constraints and did not take into account the SAR constraints.

Motivated by these observations, our work investigates an EE-aware optimization of the uplink transmit covariance matrices in an SAR-constrained MIMO system under the consideration of statistical CSI. To the authors' best knowledge, this is the first work that investigates the EE optimization with SAR constraints and exploits statistical CSI at the same time. To deal with the computationally expensive objective, we first present an asymptotic approximation of the average system SE exploiting the statistical CSI. Then, we leverage Dinkelbach's algorithm to tackle the asymptotic EE maximization problem in the form of a fractional program. Inspired by the structure of the optimal solutions to the transformed subproblems, we then customize an SAR-aware water-filling scheme to obtain an asymptotically optimal result for the EE optimization taking into account the SAR constraints. Lastly, superior performance improvements of the proposed SAR-aware EE maximization precoding approach over other baselines are illustrated in the numerical results.

\section{System Model and Problem Formulation}\label{sec:sysmod}
Consider the uplink of a single-cell MIMO system with $K$ user terminals (UTs) transmitting to one $M$-antenna BS. Denote $\bx_k \in \C^{N_k \times 1}$ as the transmit signal of UT $k$, $\forall k \in \cK = \left\{1,\ldots,K\right\}$, where $N_k$ is the number of transmit antennas. We assume that $\left\{\bx_k\right\}_{\forall k}$ are zero-mean and mutually independent, i.e., $ \expect{\bx_k \bx_{k'}^H}  = \bzero$, $\forall k' \ne k$, and denote $ \bQ_k = \expect{\bx_k \bx_k^H} \in \C^{N_k \times N_k}$ as the covariance matrix. The signal received at the BS can then be modeled as
\begin{align}
\by = \sum\nolimits_{k=1}^K { \bH_k \bx_k } + \bn \in \C^{M \times 1},
\end{align}
where $\bH_k \in \C^{ M \times N_k}$ is the uplink channel matrix of UT $k$, and $\bn \in \C^{ M \times 1} \sim \mathcal{CN}(\bzero,\sigma^2 \bI_{M})$ is the thermal noise at the BS with $\sigma^2$ being the noise power at each receive BS antenna.

The signals $\left\{ \bx_k \right\}_{\forall k}$ are commonly subject to some transmit power limitation, i.e., $\tr{\bQ_k} \le P_{\max,k}$, $\forall k$, where $P_{\max,k}$ is the given power budget. In practical systems, UTs' transmit signals should be constrained by EM exposure, which is usually measured by SAR. Herein, we adopt a theoretical SAR model to describe the spatial-averaged SAR caused by multiple antennas \cite{Radio2019Xu}, which has been widely accepted in the literature, e.g., \cite{ying2015closed,ying2017sum,zhang2020specific}. Specifically, the SAR constraints at UT $k$ are given by
\begin{align}
\SAR_{k,i} & = \expect{\tr{\bx_k^H \bR_{k,i} \bx_k}} \ntb
& = \tr{\bR_{k,i} \bQ_k} \le Q_{k,i}, \quad i = 1,\ldots,G_k,
\end{align}
where $G_k$ is the number of SAR measurements, $\bR_{k,i}$ denotes the $i$th SAR matrix imposed at UT $k$, and $Q_{k,i}$ denotes the corresponding maximum tolerable SAR value. Each SAR matrix measures one local SAR and the max of the local SARs should be smaller than the limitations for compliance testing \cite{Radio2019Xu,zhang2017analysis}. Since various EM exposure limits are set for different testing areas of the user, a device should satisfy multiple SAR constraints, which are characterized by multiple SAR matrices \cite{castellanos2019dynamic}.

As for the uplink MIMO channels, we consider the Weichselberger jointly correlated Rayleigh fading model for channel spatial correlations \cite{Gao09Statistical}. Specifically, $\bH_k$ is modeled as
\begin{align}\label{eq:Weichselberger}
\bH_{k} = \bU_{k} \beamH_{k} \bV^H_{k} ,\quad \forall k \in \cK,
\end{align}
where $\bU_{k} = \left[ \bu_1,\ldots,\bu_M \right] \in \C^{M \times M}$, $\bV_{k} = \left[ \bv_1,\ldots,\bv_{N_k} \right] \in \C^{N_k \times N_k}$ are deterministic unitary matrices, and $\beamH_{k} \in \C^{M \times N_k}$ is a random matrix with independently-distributed and zero-mean entries. In the considered statistical CSI scenario, we exploit the eigenmode coupling matrices as the statistics of $\left\{\bH_k\right\}_{\forall k}$, which are defined as \cite{Gao09Statistical}
\begin{align}
\bOmega_k = \expect { \beamH_k \odot \beamH_k^* }  \in { \R^{ M \times N_k }}, \quad \forall k \in \cK,
\end{align}
where $\odot$ denotes the Hadamard product. Denote $\bQ \triangleq \left\{ \bQ_k \right\}_{k=1}^K$, the ergodic system SE is expressed by \cite{Wen11On}
\begin{align}\label{eq:ergodic_SE}
\SE\left(\bQ\right) &= \expect { \log_2 \det  \left( \bI_M + \s \sum\nolimits_{k=1}^K \bH_k \bQ_k \bH_k^H \right) } \ntb
&\qquad [\mathrm{bits/s/Hz}],
\end{align}
where the expectations are taken over $\left\{\bH_k\right\}_{\forall k}$. For the description of EE measurement, we first introduce the system's energy consumption as follows \cite{you2020energy,you2020reconfigurable}
\begin{align}\label{eq:power_consumption}
P \left(\bQ\right) = \sum\nolimits_{k=1}^K \left( \xi_k \tr{\bQ_k} + \Pck \right) + P_{\mathrm{BS}},
\end{align}
where $\xi_k > 1$ is the reciprocal of the amplifier efficiency at UT $k$. In addition, $\Pck>0$ and $P_{\mathrm{BS}}>0$, respectively, model the $k$th UT's and BS's static hardware-dissipated power. Then, the ergodic system EE with bandwidth $W$ is defined as
\begin{align}\label{eq:ergodic_EE}
\EE\left(\bQ\right) \triangleq  W \frac{\SE\left(\bQ\right)}{P\left(\bQ\right)}\qquad [\mathrm{bits/s/Hz}].
\end{align}

In this paper, we study the SAR-aware EE maximization precoding design by optimizing $\left\{\bQ_k\right\}_{\forall k}$ and considering both the power and SAR limitations, which is cast as follows
\begin{subequations}\label{eq:EE_Q}
\begin{align}
\cP_1:\quad\underset{\bQ} \max \quad & \EE\left(\bQ\right) \label{eq:EE_Q_obj}\\
{\mathrm{s.t.}}\quad
& \tr{\bQ_k} \le P_{\max,k}, \quad \bQ_k \succeq \bzero,\quad \forall k, \\
& \tr{\bR_{k,i}\bQ_k} \le Q_{k,i}, \quad \forall k,i. \label{eq:EE_Q_sar}
\end{align}
\end{subequations}
It is in general difficult to deal with $\cP_1$ due to several reasons. First, although SE is concave, EE is a non-concave function, thus is more complicated to optimize than that of SE. Second, the computation of the objective function, i.e., the ergodic EE, requires solving high-dimensional integrals, which is generally computationally expensive. Moreover, the introduction of the SAR constraints further complicates the EE optimization problem. In the sequel, we overcome these challenges and then develop an efficient algorithm for the SAR-constrained EE optimization in $\cP_1$.

\section{SAR-Aware EE Maximization Precoding Design}
In the scenario where UTs know only the statistics of $\left\{\bH_k\right\}_{\forall k}$, it is challenging to obtain the analytical solutions of $\cP_1$ due to the expectation operations in the objective function. Although Monte-Carlo methods could be applied to estimate the expectation values in \eqref{eq:ergodic_SE}, expensive computational cost required in exhaustive channel averaging might be intolerable. Hence, for complexity reduction, we adopt an asymptotic SE expression, which relies only on the channel statistics $\left\{\bOmega_k\right\}_{\forall k}$ rather than the realizations $\left\{\bH_k\right\}_{\forall k}$. Assuming that $N_k \to \infty$, $\forall k$, $M \to \infty$ with constant ratios $c_k = N_k/M$, $\forall k$, the asymptotic SE is given by \cite{Wen11On}
\begin{align}\label{eq:DE}
& \SEde\left(\bQ\right) = \sum\nolimits_{k=1}^K \log_2 \det \left( \bI_{N_k} + \s \bGamma_k \bQ_k \right) \ntb
& \quad  + \log_2 \det \left( \bI_M + \sum\nolimits_{k=1}^K {\bPsi_k} \right) - \sum\nolimits_{k=1}^K {\bgamma_k^T \bOmega_k \bpsi_k},
\end{align}
where $\bGamma_k \in \C^{N_k \times N_k}$ and $\bPsi_k \in \C^{M \times M}$ are given by
\begin{align}\label{eq:T}
\bGamma_k & =  \bV_k \diag{ \bOmega_k^T \bgamma_k } \bV_k^H, \\ \label{eq:F}
\bPsi_k & = \bU_k \diag{ \bOmega_k \bpsi_k } \bU_k^H,
\end{align}
respectively. Moreover, given an initial point, $\bgamma_k \triangleq \left[ \gamma_{k,1},\ldots,\gamma_{k,M} \right]^T$ and $\bpsi_k \triangleq \left[\psi_{k,1},\ldots,\psi_{k,N_k} \right]^T$ are obtained by the following iterative update
\begin{align}\label{eq:gamma}
\gamma_{k,m} & = \bu_{k,m}^H \left( \bI_M + \bPsi \right)^{-1} \bu_{k,m}, \quad \forall k,m, \\ \label{eq:psi}
\psi_{k,n} & = \bv_{k,n}^H \bQ_k \left( \sigma^2 \bI_{N_k} + \bGamma_k \bQ_k \right)^{-1} \bv_{k,n},\quad \forall k, n.
\end{align}
Although \eqref{eq:DE} is derived for large-scale MIMO, it is also very accurate for small-scale MIMO and has been verified in \cite{Wen11On}.

Approximating $\SE\left(\bQ\right)$ with $\SEde\left(\bQ\right)$, an asymptotic SAR-aware EE maximization problem is then given by
\begin{subequations}\label{eq:EEde_Q}
\begin{align}
\cP_2:\quad\underset{\bQ} \max \quad & \EEde \left(\bQ\right) \triangleq \frac{\SEde\left(\bQ\right)}{ P \left(\bQ\right) }\\
{\mathrm{s.t.}}\quad
& \tr{\bQ_k} \le P_{\max,k}, \quad \bQ_k \succeq \bzero, \quad \forall k,\\
& \tr{\bR_{k,i} \bQ_k} \le Q_{k,i}, \quad \forall k,i,
\end{align}
\end{subequations}
where we omit the constant $W$ for brevity. In \cite{Wen11On}, $\SEde\left(\bQ\right)$ is shown to be strictly concave on $\left\{\bQ_k\right\}_{\forall k}$. Accordingly, the objective function of the SAR-aware EE optimization problem in \eqref{eq:EEde_Q} is quasi-concave where the numerator is concave and the denominator is affine with respect to $\left\{\bQ_k\right\}_{\forall k}$, thus can be handled via Dinkelbach's algorithm. Specifically, we tackle \eqref{eq:EEde_Q} through addressing a series of concave programs and the $\ell$th subproblem is given by
\begin{subequations}\label{eq:Dinkelbach}
\begin{align}
\cP_3:\ \bQ^{(\ell+1)} = \underset{\bQ} {\arg\max} \ & \SEde\left(\bQ\right) - \eta^{(\ell)} P \left(\bQ\right) \\
{\mathrm{s.t.}} \
& \tr{\bQ_k} \le P_{\max,k}, \ \bQ_k \succeq \bzero,\ \forall k,\\
& \tr{\bR_{k,i} \bQ_k} \le Q_{k,i}, \ \forall k,i,
\end{align}
\end{subequations}
where
\begin{align}\label{eq:eta_update}
\eta^{(\ell)} =  \frac{\SEde\left(\bQ^{(\ell)}\right)} { P \left(\bQ^{(\ell)}\right)}.
\end{align}

Exploiting the convergence properties described in \cite[Proposition 3.2]{zappone2015energy}, the solutions of Dinkelbach's transformed subproblems \eqref{eq:Dinkelbach}, i.e., $\left\{\bQ^{(\ell)}\right\}_{\ell}$, are guaranteed to converge to the global optimum of the asymptotic SAR-aware EE optimization in \eqref{eq:EEde_Q}. The remaining difficulty arises in devising a computationally-efficient way for solving \eqref{eq:Dinkelbach}.

Our approach is to tackle \eqref{eq:Dinkelbach} by solving its dual problem. To this end, we define the Lagrangian of Dinkelbach's transformed problem in \eqref{eq:Dinkelbach} as
\begin{align}\label{eq:Lagrangian1}
& \quad \cL\left( \bQ, \left\{\mu_k\right\}_{\forall k}, \left\{\beta_{k,i}\right\}_{\forall (k,i)} \right)\ntb
& = \ \SEde\left(\bQ\right) - \eta^{(\ell)} P \left(\bQ\right) - \sum\nolimits_{k=1}^K \mu_k \left( \tr{\bQ_k} -  P_{\max,k} \right) \ntb
& \quad - \sum\nolimits_{k=1}^K \sum\nolimits_{i=1}^{G_k} \beta_{k,i} \left( \tr{\bR_{k,i} \bQ_k} - Q_{k,i} \right),
\end{align}
where $\left\{\mu_k\right\}_{\forall k}$ and $\left\{\beta_{k,i}\right\}_{\forall (k,i)}$ are the Lagrange multipliers for the power and SAR constraints, respectively. Then, the dual problem of \eqref{eq:Dinkelbach} is given by
\begin{align}\label{eq:dual}
\underset{ \begin{subarray}{c} \mu_k \ge 0, \forall k \\ \beta_{k,i} \ge 0, \forall k,i \end{subarray} } \min \ \underset{\bQ_k \succeq \bzero, \forall k} \max \ \cL\left( \bQ, \left\{\mu_k\right\}_{\forall k}, \left\{\beta_{k,i}\right\}_{\forall (k,i)} \right).
\end{align}
As shown in \cite{Wen11On}, the derivative of $\SEde\left(\bQ\right)$ over each matrix-valued variable $\bQ_k$ is given by
\begin{align}\label{eq:derivative}
\frac{\partial \SEde\left(\bQ\right) }{\partial \bQ_k} & = \frac{\partial \log_2 \det  \left( \bI_{N_k} + \s \bGamma_k \bQ_k \right) }{\partial \bQ_k} \\
& = \frac{1}{\ln 2}\left( \sigma^2 \bI_{N_k} + \bGamma_k \bQ_k \right)^{-1} \bGamma_k, \quad \forall k.
\end{align}
Accordingly, with given $\bgamma \triangleq \left\{\gamma_{k,m}\right\}_{\forall k,m}$ and $\bpsi \triangleq \left\{\psi_{k,n}\right\}_{\forall k,n}$, the inner optimization in \eqref{eq:dual}, i.e., the maximization of $\cL$ with respect to $\left\{ \bQ_k \right\}_{\forall k}$ is equivalent to
\begin{align}\label{eq:Lagrangian2}
\underset{\bQ_k\succeq \bzero} \max \ \log_2\det \left( \bI_{N_k} + \s \bGamma_k \bQ_k \right) - \tr{\bK_k \bQ_k} , \quad \forall k,
\end{align}
where
\begin{align}\label{eq:Kk}
\bK_k = \left( \eta^{(\ell)} \xi_k + \mu_{k} \right) \bI_{N_k} + \sum\nolimits_{i=1}^{G_k} \beta_{k,i} \bR_{k,i}.
\end{align}

The equivalent Lagrangians in \eqref{eq:Lagrangian2} demonstrate the structure of the optimal solutions to the asymptotic SAR-aware EE optimization in \eqref{eq:Dinkelbach}, thus enabling us to conceive a computationally-efficient algorithm for \eqref{eq:Dinkelbach}. Specifically, the optimal results of \eqref{eq:Dinkelbach} can be determined in the following proposition.
\begin{prop}\label{prop:opitmal_EVD}
Denote the optimal dual variables as $\left\{\mu_{k,\opt}\right\}_{\forall k}$ and $\left\{\beta_{k,i,\opt}\right\}_{\forall k,i}$. Accordingly, we denote $\bK_{k,\opt} = \left( \eta^{(\ell)} \xi_k + \mu_{k,\opt} \right) \bI_{N_k} + \sum\nolimits_{i=1}^{G_k} \beta_{k,i,\opt} \bR_{k,i}$ and the eigenvalue decomposition of $\bK_{k,\opt}^{-1/2} \bGamma_k \bK_{k,\opt}^{-1/2}$ is given by
\begin{align}\label{eq:EVD_K}
\bK_{k,\opt}^{-1/2} \bGamma_k \bK_{k,\opt}^{-1/2} = \bU_k \bSigma_k \bU_k^H,
\end{align}
where $\bSigma_k = \diag{p_{k,1},\ldots,p_{k,N_k}}$ with $p_{k,1} \ge p_{k,2} \ge \ldots \ge p_{k,N_k} \ge 0$. Then, the optimal solutions of \eqref{eq:Dinkelbach} can be expressed as
\begin{align}\label{eq:EVD_optimalQ}
\bQ_{k,\opt} = \bK_{k,\opt}^{-1/2} \bU_k \bLambda_k \bU_k^H \bK_{k,\opt}^{-1/2}, \quad \forall k,
\end{align}
where
\begin{align}\label{eq:optimal_Lambda}
\bLambda_k = \diag{ \left(1 - \sigma^2/p_{k,1}\right)^+,\ldots,\left(1 - \sigma^2/p_{k,N_k}\right)^+},
\end{align}
with the notation $[x]^+ = \max (x,0)$. In addition, $\left\{\mu_{k,\opt}\right\}_{\forall k}$ and $\left\{\beta_{k,i,\opt}\right\}_{\forall k,i}$ are chosen to meet the Karush-Kuhn-Tucker (KKT) optimality conditions given by
\begin{align}\label{eq:KKT_power}
& \mu_{k,\opt} \left( \tr{\bQ_{k,\opt}} -  P_{\max,k} \right) = 0,\ntb
& \mu_{k,\opt} \ge 0, \quad \tr{\bQ_{k,\opt}} \le P_{\max,k},\quad \forall k,
\end{align}
and
\begin{align}\label{eq:KKT_SAR}
& \beta_{k,i,\opt} \left( \tr{\bR_{k,i} \bQ_{k,\opt}} - Q_{k,i} \right) = 0, \ntb
& \beta_{k,i,\opt} \ge 0, \quad \tr{\bR_{k,i}\bQ_{k,\opt}} \le Q_{k,i},\quad \forall k,i.
\end{align}
\end{prop}

\begin{remark}In fact, it can be shown that when SAR constraints are not considered, the eigenvectors of $\bQ_k$ admit closed-form deterministic solutions and therefore only the eigenvalues, i.e., $\bLambda_k$, have to be optimized. However, the solution in \eqref{eq:EVD_optimalQ} shows that both the eigenvectors and the eigenvalues of $\bQ_k$ have to be optimized, which indicates that the EE optimization with SAR constraints is more challenging to deal with than the one without SAR constraints.
\end{remark}

The proof of \propref{prop:opitmal_EVD} can be derived via a similar method as that of \cite[Theorem 3.6]{ying2015closed}, which is omitted here due to space limitation. The optimal $\left\{\mu_{k,\opt}\right\}_{\forall k}$ and  $\left\{\beta_{k,i,\opt}\right\}_{\forall k,i}$ are determined by the dual problem in \eqref{eq:dual}, which is convex and therefore its solution can be found by standard convex optimization approaches, e.g., the sub-gradient method \cite{Boyd04Convex}.

\propref{prop:opitmal_EVD} reveals that each $\bQ_{k,\opt}$ follows an SAR-aware EE maximization water-filling principle over $\bK_{k,\opt}^{-1/2} \bGamma_k \bK_{k,\opt}^{-1/2}$. Meanwhile, $\left\{\bGamma_k\right\}_{\forall k}$ are in turn dependent on $\left\{\bQ_{k,\opt}\right\}_{\forall k}$. Therefore, alternating optimization (AO) is applied to optimize between $\left\{\bQ_{k,\opt}\right\}_{\forall k}$ and $\left\{\bGamma_k\right\}_{\forall k}$, where we update $\left\{\bQ_{k,\opt}\right\}_{\forall k}$ by \propref{prop:opitmal_EVD} with given $\left\{\bGamma_k\right\}_{\forall k}$, and then calculate $\left( \bgamma,\bpsi \right)$ by \eqref{eq:gamma} and \eqref{eq:psi} with the updated $\left\{\bQ_{k,\opt}\right\}_{\forall k}$ to update $\left\{\bGamma_k\right\}_{\forall k}$. Based on the iterative process, an asymptotic optimization algorithm for the SAR-aware EE optimization in \eqref{eq:EE_Q} is detailed in \alref{alg:alg1}.

Now, we discuss the convergence of the proposed SAR-aware energy-efficient precoding approach in \alref{alg:alg1}, which is based on Dinkelbach's transformation. During each step, we solve the Dinkelbach's subproblem in \eqref{eq:Dinkelbach} based on \propref{prop:opitmal_EVD}. Since \eqref{eq:Dinkelbach} is a concave program and the results in \propref{prop:opitmal_EVD} are in general derived by solving the KKT optimality conditions of its dual problem in \eqref{eq:dual}, the iterative modified water-filling scheme in \alref{alg:alg1} based on \propref{prop:opitmal_EVD} converges to the globally optimal solution of \eqref{eq:Dinkelbach} \cite{Boyd04Convex}. Then, exploiting the convergence properties of Dinkelbach's transformation, the solution sequence generated by \eqref{eq:Dinkelbach} converges to the global optimum of the asymptotic SAR-aware EE optimization in \eqref{eq:EEde_Q} \cite{zappone2015energy}. Thus, \propref{prop:opitmal_EVD} is guaranteed to converge.

Then, we discuss the complexity of \alref{alg:alg1} as follows. The main structure of \alref{alg:alg1} is based on Dinkelbach's transformation, which is assumed to require a total of $I_{\mathrm{D}}$ iterations. During each iteration, we assume that the convergence of the dual variables require $I_{\mathrm{d}}$ iterations. Each time given the dual variables, we utilize the AO method to update $\bQ_k$ and $\bGamma_k$ in an iterative manner, which is assumed to require $I_{\mathrm{AO}}$ iterations. Owning to the fast convergence rate of $\left( \bgamma,\bpsi \right)$ \cite{Couillet11Random}, as well as the low complexity of calculating $\bGamma_k$, the per-iteration complexity of AO is mainly dominated by that of updating $\bQ_k$. The major complexity of updating $\bQ_k$ lies in the complexity of the eigenvalue decomposition in \eqref{eq:EVD_K}, which is $\cO\left(N_k^3\right)$. Moreover, the complexity of updating the dual variables by minimizing $\cL$ in step 11 is estimated as $\cO\left( \left(1+G_k\right)^p \right)$, where $1 \le p \le 4$ for standard convex program solutions \cite{huang2019reconfigurable}. Therefore, the complexity of \alref{alg:alg1} is estimated as $\cO\left( I_{\mathrm{D}} I_{\mathrm{d}} \left( \sum\nolimits_k \left( I_{\mathrm{AO}} N_k^3  + \left(1+G_k\right)^p \right) \right) \right)$. Note that the values of $I_{\mathrm{D}}$, $I_{\mathrm{d}}$, and $I_{\mathrm{AO}}$ depend on the preset thresholds.

\begin{algorithm}[h]
\caption{SAR-Aware Energy-Efficient Precoding Algorithm.}
\label{alg:alg1}
\begin{algorithmic}[1]
\State Initialize feasible $\bQ_k^{(0)}$, $\forall k$, $\eta^{(0)} = 0$, threshold $\varepsilon$, and set iteration index $\ell=0$.
\Repeat
\State Initialize $\mu_k \ge 0$, $\forall k$, and $\beta_{k,i} \ge 0$, $\forall k,i$.
\Repeat
\State Initialize $\bX_k^{0} = \bQ_k^{(\ell)}$, $\forall k$, and set iteration index $t=0$.
\Repeat
\State Update $(\bgamma^{t},\bpsi^{t})$ by \eqref{eq:gamma} and \eqref{eq:psi} with $\bX_k^{t}$, $\forall k$.
\State Update $\bGamma_k^{t}$, $\forall k$, by \eqref{eq:T} with $(\bgamma^{t},\bpsi^{t})$.
\State Update $\bX_k^{t+1}$, $\forall k$, based on \propref{prop:opitmal_EVD} with $\eta^{(\ell)}$, $\bGamma_k^{t}$, $\forall k$, and given dual variables.
\Until{$\cL$ converges with given dual variables.}
\State Update $\mu_k$, $\forall k$, and $\beta_{k,i}$, $\forall k,i$ by minimizing $\cL$.
\Until{$\left\{\mu_k\right\}_{\forall k}$ and  $\left\{\beta_{k,i}\right\}_{\forall k,i}$ converge.}
\State Set $\bQ_k^{(\ell+1)} = \bX_k^{t+1}$, $\forall k$.
\State Update $\eta^{(\ell+1)}$ by \eqref{eq:eta_update} with $\bQ_k^{(\ell+1)}$, $\forall k$.
\State Set $\ell = \ell + 1$.
\Until{$\left| \eta^{(\ell)} - \eta^{(\ell-1)} \right|\le \varepsilon$.}
\end{algorithmic}
\end{algorithm}

\begin{remark}
Note that although \alref{alg:alg1} is custom-designed for the EE maximization, it can be applied for the SE maximization with slight modification.
In particular, by simply setting $\xi_k = 0$, $\forall k$, maximizing the EE is equivalent to maximizing its numerator, i.e., the SE. Then, \alref{alg:alg1} can be specialized into the SE maximization, where only one iteration of Dinkelbach's transformation is needed.
\end{remark}

\section{Numerical Results}\label{sec:numerical}
In this section, we present computer simulations to illustrate the effectiveness of the proposed SAR-constrained EE maximization transmission in multiuser MIMO uplink with statistical CSI. The 3GPP propagation environment is applied for channel generation \cite{Salo05MATLAB}. The major simulation parameters are set as follows \cite{you2020spectral,you2020reconfigurable,ying2017sum}: transmission bandwidth: $W = 10$ MHz, large scale fading: $-120$ dB, noise variance: $\sigma^2 = -96$ dBm, number of UTs: $K = 4$, number of antennas at each UT and the BS: $N_k = 4$, $\forall k$, and $M = 8$, amplifier inefficiency: $\xi_k = 1/0.2$, $\forall k$, hardware-dissipated power at each UT and the BS: $P_{\mathrm{c},k} = 30$ dBm, $\forall k$, and $P_{\mathrm{BS}} = 40$ dBm. The SAR constraints are characterized by the SAR matrices in \eqref{eq:SAR matrix} taken from \cite{castellanos2019dynamic,ying2017sum},\footnote{Note that the SAR matrices in \eqref{eq:SAR matrix} are manually generated to evaluate the proposed approach \cite{castellanos2019dynamic,ying2017sum}. A more realistic set of local SAR matrices based on a proper SAR experiment will be considered in the future work.} shown at the top of this page,
\begin{figure*}
\begin{align}\label{eq:SAR matrix}
\bR_1 = \left[ \begin{matrix} 8 & -6 \jmath & -2.1 & 0 \\ 6 \jmath & 8 & -6 \jmath & -2.1 \\ -2.1 & 6 \jmath & 8 & -6 \jmath \\ 0 &  -2.1 & 6 \jmath & 8 \end{matrix} \right], \
\bR_2 = \left[ \begin{matrix} 3.94 & -2.65 - 2.53 \jmath & -0.01 + 3.46 \jmath & 0.60 - 0.10 \jmath \\ -2.65 + 2.53 \jmath & 4.57 & -2.30 - 2.80 \jmath & -0.99 - 0.07 \jmath \\ -0.01 - 3.46 \jmath & -2.30 + 2.80 \jmath & 4.97 & -1.22 - 2.04 \jmath \\ 0.60 + 0.10 \jmath & -0.99 + 0.07 \jmath & -1.22 + 2.04 \jmath & 3.18 \end{matrix} \right].
\end{align}
\hrule
\end{figure*}
where $\bR_{k,1} = \bR_1$, $\bR_{k,2} = \bR_2$, $\forall k$, and $\jmath = \sqrt{-1}$ is the imaginary unit. The SAR constraints are set as $Q_1 = 1.0$ W/kg for $\bR_1$ and $Q_2 = 0.8$ W/kg for $\bR_2$ \cite{ying2017sum}. Moreover, we consider the same power for all UTs, i.e., $P_{\max,k} = P_{\max}$, $\forall k$.

\begin{figure}[!h]
\includegraphics[width=0.45\textwidth]{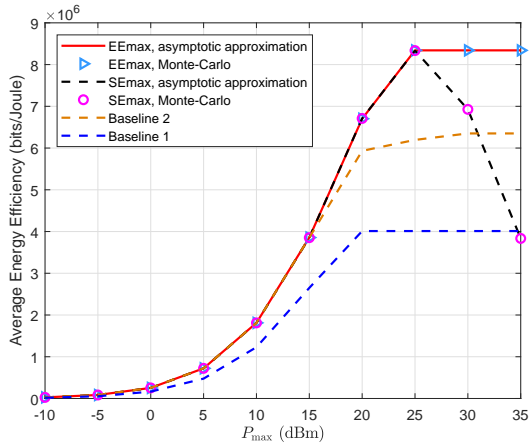}
\centering
\caption{Comparison of the proposed approach with the SE maximization approach and two back-off baselines.}
\label{fig:baselines}
\end{figure}

To validate the benefits of the proposed EE maximization (EEmax) approach, \figref{fig:baselines} compares it with the SE maximization (SEmax) one, which is a special case of \alref{alg:alg1} (by setting $\xi_k = 0$, $\forall k$). As depicted, for small $P_{\max}$, the performance of the EE maximization design is aligned with that of the SE maximization, which reveals that maximizing EE and SE are virtually the same at such power constraint levels under the SAR constraints. Whereas, for large $P_{\max}$, the two designs with different criteria yield quite different performances. Specifically, the EE value achieved by the EE maximization design becomes a constant while that achieved by the SE maximization design drops quickly as $P_{\max}$ grows. In fact, there exists an optimal transmit power that maximizes the system EE such that any allocated power exceeding the threshold is redundant and will only decrease the system EE. For EE maximization, the actual allocated transmit power would remain at the optimal level for achieving the maximum EE when $P_{\max}$ is larger than the threshold. However, the SE maximization design always exhausts all the full available transmit power to increase SE. Therefore, when $P_{\max}$ is larger than the threshold, the SE maximization design still allocates the full power budget, which leads to a smaller EE than that of the EE maximization design. To verify the accuracy of the proposed asymptotic approximation, we also compare the results of \alref{alg:alg1} with those obtained by Monte-Carlo simulations, in which the expectation values are evaluated through exhaustive channel averaging. \figref{fig:baselines} showcases an exact agreement between the asymptotic results and the Monte-Carlo results, which confirms the high accuracy of the derived asymptotic method adopted in \alref{alg:alg1}.

Traditionally, the power back-off approach has been widely adopted to deal with SAR constraints \cite{ying2015closed}. For further evaluation of the proposed SAR-constrained EE maximization approach, we compare it with two relevant baselines based on the idea of power back-off:
\begin{itemize}
\item \textbf{Baseline 1 (worst-case power back-off approach):} In this case, SAR mitigation is achieved by power reduction. Specifically, each transmit covariance matrix is designed under only the power constraint, i.e.,
    \begin{align}\label{eq:baseline1}
    \bQ_k^{\mathrm{b1}} = \underset{\tr{\bQ_k} \le \alpha_1 P_{\max}} \max \EEde \left(\bQ\right), \quad \forall k,
    \end{align}
    where the power budget is reduced as $\alpha_{1} P_{\max}$ with $\alpha_1 \le 1$ denoting the power back-off factor and given by
    \begin{align}\label{eq:alpha}
    \alpha_1 = \min \left\{ 1, \frac{Q}{\mathrm{SAR}^\mathrm{worst}_1}, \ldots, \frac{Q}{\mathrm{SAR}^\mathrm{worst}_K} \right\}.
    \end{align}
    Notice that $\mathrm{SAR}^\mathrm{worst}_k$ is called the worst-case SAR at UT $k$, which is defined as
    \begin{align}\label{eq:SAR worst}
    \mathrm{SAR}^\mathrm{worst}_k = \underset{\tr{\bQ_k} \le P_{\max}} \max \tr{\bR \bQ_k}, \quad \forall k.
    \end{align}
\item \textbf{Baseline 2 (adaptive power back-off approach):} In this case, the transmit covariance matrices are first optimized without considering the SAR constraints, i.e.,
    \begin{align}\label{eq:baseline2}
    \bQ_k^{0} = \underset{\tr{\bQ_k} \le P_{\max}} \max \EEde \left(\bQ\right), \quad \forall k.
    \end{align}
    Then, the power reduction factor $\alpha_2 \le 1$ is calculated as
    \begin{align}\label{eq:alpha2}
    \alpha_2 = \min \left\{ 1, \frac{Q}{\tr{\bR \bQ_1^{0}}}, \ldots, \frac{Q}{\tr{\bR \bQ_K^{0}}} \right\}.
    \end{align}
    At last, the final results are given by $\bQ_k^{\mathrm{b2}} = \alpha_2 \bQ_k^{0}$, $\forall k$.
\end{itemize}

\figref{fig:baselines} sketches the performance of the two baselines. We observe that both the proposed and the adaptive back-off approaches realize significant EE improvements over the worst-case back-off one. The proposed SAR-aware EE maximization approach performs almost identically to the adaptive back-off one when $P_{\max}$ is small and yields better EE performance than the back-off one when $P_{\max}$ is large. This is because when $P_{\max}$ is small, the SAR constraint is always satisfied even if full transmit power budget is used. However, for high transmit power levels, the power back-off methods \cite{ying2015closed,Hochwald2012minimizing} need to reduce the transmit power in all channels by a same factor, which results in a rapid decrease in the system EE. This shows that our considered SAR-constrained optimization could not only ensure the compliance of SAR, but also achieve better EE performance than that of the conventional power back-off schemes.

\begin{figure}[!h]
\includegraphics[width=0.45\textwidth]{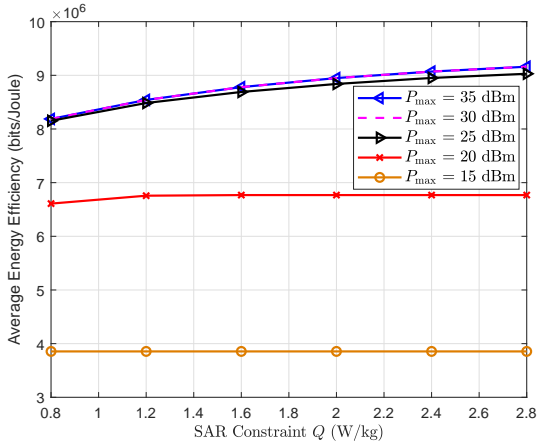}
\centering
\caption{Average EE versus SAR constraints under different maximum power budgets.}
\label{fig:SAR_constraint}
\end{figure}

To show the impact of different values of SAR constraints on the EE performance, we plot the system EE versus SAR constraints under different power constraints in \figref{fig:SAR_constraint}, where we assume $Q_1 = Q_2 = Q$. For the case of $P_{\max}=15$ dBm, EE remains a constant as the SAR constraint $Q$ changes. This is because when $P_{\max}$ is low, the maximum EE is limited by the stringent power constraint, so that increasing the SAR constraint $Q$ has no impact on EE. For the case of $P_{\max}=20$ dBm, EE first increases with $Q$ and then approaches a constant because EE is first limited by the SAR constraint and then limited by the power constraint. For cases of $P_{\max} \ge 25$ dBm, EE increases with increasing $Q$, which indicates that SAR constraints have a significant impact on EE, especially when the power budget is sufficiently high. Moreover, the results of $P_{\max}=30$ dBm are identical to those of $P_{\max}=35$ dBm since EE is saturated in high power budget regions, as has been shown in \figref{fig:baselines}.

\section{Conclusion}\label{sec:conclusion}
In this paper, we studied the SAR-aware energy-efficient transmission scheme for SAR-constrained multiuser MIMO uplink. Considering the statistical CSI scenario, asymptotic approximation results were derived which facilitates a low-complexity design of the UT's transmit covariance matrices. In particular, the asymptotic SAR-aware EE maximization problem was converted into a series of convex programs via using Dinkelbach's algorithm. We showed that the optimal solutions of these transformed subproblems have the interpretation of modified water-filling. Inspired by this insight, we proposed an iterative SAR-aware water-filling approach for the considered SAR-constrained EE maximization precoding problem. Computer simulations showcased that the developed SAR-aware EE maximization scheme introduced considerable EE improvements over the conventional power back-off methods, especially in cases with high power budgets.


\end{document}